\documentclass[12pt]{iopart}
\usepackage{graphicx}

\begin{document}
\title{{\bf Peano modes at the D=2 delocalization transition}}

\author{V. G. Benza}
\address{Dipartimento di Scienza e Alta Tecnologia, 
Universita' dell' Insubria,20064 Como, Italy} 
\ead{vincenzo.benza@uninsubria.it}
\begin{abstract}
We study the elasticity of the plane-filling Peano chain in terms of a spring network,
modeling a flat and thin elastic strip, smoothly curved 
so as to follow the chain pattern.\\ 
This network sustains normal modes, named Peano modes;
in the same way as acoustic waves in crystals are sustained by the matching of translations with rotations,
these modes arise when discrete dilatations match with $\pi/4$ rotations.
The pattern shares this eightfold symmetry with the octagonal quasicrystal, but is much looser and prone
to disruption; its relaxation marks the saddle point separating the solid from the liquid. 
In terms of the equivalent quantum mechanical problem
the above corresponds to the D=2 quantum delocalization transition.\\
Some experimental observations, regarding a priori uncorrelated systems such as
colloids, bilayer water and chromosomal DNA come together, at least qualitatively, within the viewpoint proposed here.
With reference to the D=2 colloids the model accounts for the bosonic peak found in their phononic spectrum 
and for the relaxation of the excitations.
The latter typically evolve towards patterns of increasingly large size, as in the inverse energy cascade of D=2 turbulence.
As for bilayer water, it undergoes a liquid to quasicrystal transition
surprisingly similar to the phase transition found here.
Finally, considering the chromosomal DNA, the model unveils the mixed nature of its compact conformations,  
lying at the boundary between a liquid and a solid, similar to colloids and quasicrystals. 
\end{abstract}
\maketitle

\section{\bf Introduction}

In this work we study the harmonic spectrum and the time relaxation of the planar Peano-Hilbert
curve $PH_{2}$. 
The curves $PH_{D}$ that fill D-dimensional spaces are used
in ordering multidimensional data \cite{ziv,weissmann},
as well as in manufacturing miniature antennas \cite{engheta}.
In the domain of polymer physics the curve $PH_{3}$ has  been proposed    
as a possible final state of a scale-invariant collapse \cite{grosberg}.
Quite remarkably, given the high degree of ordering of $PH_{3}$, its pattern has been
found to fit with the spatial organization of the chromosomal DNA in humans
over a rather large range of scales \cite{lieberman}.
This resulted from detecting the physical contact frequencies between pairs of
genomic loci, using chromosome conformation capture methods \cite{dekker}.
Regarding DNA dynamics, significant data emerged from experiments
on bacterial (E.coli) chromosomes \cite{weber,reyes-lamothe}.
It was found that DNA loci undergo anomalous diffusion with a rather sharp
distribution in the exponents, where the dominant behavior $(t^{0.4})$
can be connected with a fractal dimension $D_{f} \leq 3$ \cite{benza}.
A physical description of the chromosomal conformation could be
of interest for cell biologists and should be based on quantitative informations regarding   
the chain topology, the distribution of crosslinks, the nature of interchain interactions;
in addition, this looked-for description should account for the complex multicomponent 
medium of the cell.
Originally motivated by this phenomenology, the present paper is limited to analizing
the harmonic properties of the planar curve $PH_{2}$.
We consider a quasi one-dimensional elastic strip, modeled
as a mechanical network running along $PH_{2}$,
where deviations from the exact pattern carry a harmonic energy.  
The elastic couplings are anisotropic since they vary according with the local geometry, 
so that the resulting mechanical network  turns out to be rigid:
apart from uniform translations it does not allow for zero-energy modes.
In particular, rigid rotations or uniform dilatations of any portion of the network have an energetic cost.
The model does not account for the interaction among physically adjacent sites when they are distant along the chain: 
our aim is to describe waves propagating along meandering one-dimensional patterns,
not necessarily having the nature of covalent bonds.
This description could as well apply to electromagnetic waves, 
as it is known that the D=2 elasticity has strong similarities with the D=2 electromagnetism:
the elastic deformations can be decomposed into irrotational and solenoidal components, where 
dilatation $(\mathcal{D})$ and shear $(\mathcal{S})$ have the roles of electric and magnetic field respectively. 
The main difference between the two theories regards the wave propagation properties, because
dilatation pays an higher energy with respect to shear, and as a consequence in the bulk 
the longitudinal $(\mathcal{D})$ field propagates with an higher velocity with respect to the transverse  
$(\mathcal{S})$ field.
The two distinct waves couple at the medium's boundary, 
where they form the  Rayleigh waves, which are characterized by nontrivial boundary-dependent
dispersion laws.
This dependence is exploited to extract information during earthquakes, 
and in the case of electromagnetism it can be used both to extract and to manipulate information,
as done, e.g., by scattering electromagnetic waves through photonic crystals \cite{niu}
or by engineering e.m. wavefronts with metamaterials \cite{alu}.
In the elastic model in question here the particular ordering enforced by the Peano pattern, which is built by
 inflating quadrupolar partitions of planar regions, produces non-dispersive Rayleigh waves, named Peano waves.\\
In order to make clear the physical interpretation of the network, we recall 
how contact interactions among rigid disks undergo harmonic modeling, 
a procedure going under the name of fluctuation method \cite{salsburg}.  
In that framework the notion of contact (harmonic) network \cite{farago,chakra} can be given the following precise meaning 
\cite{brito07,brito06}:
'In a metastable state of  a hard sphere system, two particles are considered to be in contact 
when they collide during some time interval $\Delta t$ much smaller than the
relaxation time of the structure, where metastability is lost, 
and much larger than the (thermal) collision time'.
If the Peano pattern is to be associated with a linear polymer, 
the model describes the polymer's fluctuations within a steep effective potential.
Notice that the potential has its own time evolution, but on the larger scale of the structural relaxation.
At shorter times the line of minima of the potential
follows the topology assigned by the ordered sequence of bendings of the undeformed pattern.
Within this polymeric interpretation, if one were to
 add to the picture the contact couplings among adjacent but nonsubsequent sites,
one would end up with an elastic network  describing a textured membrane.
The phononic spectrum of such enlarged model would then display in the density of states, 
in addition to the contributions described in this work,
the behavior $\rho(\lambda)_{D=2} \approx \lambda$, as expected from standard D=2 Debye spectra.
It is worth mentioning that some experiments on two-dimensional colloids did in fact display
an anomalously high density of states at low frequencies \cite{yodh,kurchan}, superimposed to
the D=2 Debye behavior. As a matter of fact we show here that
the connection is not limited to the so-called bosonic peak of colloids, 
but involves their space-time behavior as well.
In analogy with quasicrystals, whose phononic spectra are marked by self-similarity \cite{kohmoto,ashraff,luck},
we exploit the recursive nature of the curve and consider the sequence of its periodic approximants
having the form of closed chains on the square lattice.
At the lattice sites
the harmonic operator acts on the two-dimensional space of planar deformations; it
has a $( 2 \times 2)$ block tridiagonal structure since it involves, at each site, the two nearest neigbours 
along the chain.
For the sake of brevity we will call this operator disordered laplacian, in spite of it being neither scalar nor disordered.
In fact the operator is deterministic, as the sequence of the different vertices that assigns the chain pattern.\\
We have found that in the case of  $PH_{2}$ the spectral problem for the disordered laplacian 
translates  into a finite-dimensional trace map, which can be treated by standard methods.  
The phononic spectrum of the closed curve shows 
a multiplicity of power-law singularities in the density of states,
with exponents covering a whole interval (see Figs.1 and 2).
The physics implied by such an exotic spectrum emerges from the analysis
of the relaxation process, described by the overdamped  
Langevin equation associated with the disordered laplacian.
The result is that every exponent identifies 
the scaling behavior (with respect to length and time) of a specific class of fluctuations.\\
We then found that the anisotropy of the elastic coupling 
induces a phase separation (see Fig.3);
the two phases are respectively dominated by dilatation $(\mathcal{D})$ and shear $(\mathcal{S})$, and can be
identified with solid and liquid.
The critical point where the two phases coalesce corresponds to the isotropic limit;
quite remarkably the interference of  $(\mathcal{D})$ and  $(\mathcal{S})$ generates not only anomalous fluctuations, 
but also  phononic modes, the Peano modes. 
During the relaxation process 
the energy flows from localized to extended structures,  as in the D=2 turbulence \cite{tabeling}
and as observed in the mentioned experiments on two-dimensional
colloids \cite{yodh,kurchan}.
From the perspective of the quantum-mechanics determined by the disordered laplacian, 
the isotropic limit corresponds to the critical point of the D=2 quantum delocalization transition. 
In other words the Peano chain identifies a two-dimensional extension of the D=1 Aubry-Andre' class \cite{aubryandre}.
A comprehensive exploration of the critical point requires an analysis of the trace map; 
here we limit ourselves to some preliminaries regarding its symmetry properties.
As discussed in Section 6  we find an eightfold symmetry, revealing that $PH_{2}$ is close to the octagonal quasicrystal, 
but with a definitely lower connectivity. Interestingly, 
a liquid to dodecagonal quasicrystal transition has been reported for bilayer water \cite{molinero} in a similar regime.
Some preliminaries regarding the properties of the Peano chain are reported
in the Appendix where, more importantly, we derive a substitution rule for closed chains.
In Section 2 we introduce the network model and the disordered laplacian; we then formulate
the eigenvalue problem for this laplacian in terms of a trace map. 
In Section 3 we analize the singularities of the spectral density of states by means of the so-called 
thermodynamic formalism of multifractals.
Readers uninterested in the spectral properties can  skip Sections 2 and 3 and  
go over to Section 4, where we examine the fluctuations.
The anomalous diffusion exponents  and fractal dimensions are explicitly obtained
for the planar case and estimated by means of a mean-field argument valid for chains of arbitrary
dimension D. The exponents obtained with $D=3$ appear to be in agreement 
with measures of anomalous diffusion performed on bacterial nucleoids \cite{weber,reyes-lamothe}.\\ 
Section 5 is centered on the solid-liquid phase transition and on the properties of the relaxation process. 
In Section 6 we discuss the symmetries of the system at the critical point and illustrate
how they relate with the Peano modes.

\section{\bf Elastic network and trace map}

We consider the vector field of deformations of the plane-filling Peano-Hilbert chain $PH_{2}$.
An example of the patterns described by $PH_{2}$ is given in Fig.2A; as it is 
shown in the Appendix, the closed chain can be constructed by means of an iterative procedure defined over the square lattice.
One starts with the approximant of order $k=1$, taken as the square path, and inflates the paths
with suitable rules.
The approximant of order $k$ is a closed chain that covers with no intersection all sites of the  $(2^{k} \times 2^{k}),\, (k=1,2,...)$ lattice. 
Upon assigning an orientation to the path the lattice sites can be labeled
with the chain coordinate $n, \,(n=1,2,...4^{k})$, and the vector field of deformations can be written as 
$\vec{x}(n) \equiv (x(n), y(n))$.  
This labeling is appropriate in our context since  
we assume that the excitations are transmitted only along the 
Peano pattern and that their elastic energy depends  on its local geometry. 
The dynamics studied here involves
the two-dimensional spaces of deformations at the sites $(n-1,\, n,\, n+1)$. 
The normal modes of the model solve the following equation:
\begin{equation}
\lambda \,\, \hat{I} \vec{x}(n)= 
\hat{\Gamma}^{(in)}_{n}  \vec{x}(n-1) + \hat{ \Gamma}^{(out)}_{n} \vec{x}(n+1) + \hat{V}_{n} \vec{x}(n),
\end{equation}
where $\hat{I}$ is the identity and $\hat{\Gamma}^{(in)},\, \hat{\Gamma}^{(out)}$ are the $2 \times 2$ matrices describing the contributions
from the (incoming/ outgoing) links attached to the $n$-th site.
The general form of the $\Gamma$'s, when including an interaction that mixes the $x$ and $y$
components, is given by:
\begin{equation}
\Gamma_{x,x}= k_{x},\,\,\,\, \Gamma_{x,y}= \gamma
\end{equation}
\[
\Gamma_{y,x}=\gamma, \,\,\,\Gamma_{y,y}=k_{y}
\]
The operator $\hat{V}$ is the contribution from the vertex that is determined by requiring the invariance under uniform translations:
\begin{equation}
 \hat{V}_{n} + \hat{\Gamma}^{(in)}_{n}  + \hat{ \Gamma}^{(out)}_{n} = \hat{0},
\end{equation}
this condition thus guarantees that the eigenvalue $\lambda =0$ is in the spectrum at its upper bound.
The network is anisotropic because we assign an higher energy to the deviations
when they are orthogonal to the local alignement of the pattern.
If we label with a $t$ (transverse) and with an $l$ (longitudinal) the corresponding elastic constants we thus 
have $k_{t},\, k_{l},\,\,\, k_{t} > k_{l}$.
In correspondence with an horizontal link the diagonal entries of the matrix $\Gamma$
are $k_{x}=k_{l}, \, k_{y}=k_{t}$; the off-diagonal entries $\gamma$ can describe, 
in the polymeric interpretation, the elasticity of the chain bendings.
One can notice that even in the absence of this contribution the model has no soft (zero energy) modes
other than the uniform one. 
On the r.h.s. of the eigenvalue equation one has an operator
with the appearance of a matrix-valued inhomogeneous laplacian
that acts on the two-dimensional vector field of deviations.
For the sake of brevity let us name it disordered laplacian  $\Delta_{(PH)}$, with the proviso
that its inhomogeneity is actually deterministic since it is determined by the Peano pattern.
When the off-diagonal contribution is disregarded $(\gamma = 0)$ the  x and y components are decoupled,
and one is left with two scalar disordered laplacians $\Delta_{x}$ and $\Delta_{y}$:
\begin{equation}
\Delta_{(PH)} \equiv Diag(\Delta_{x}(k_{t}, k_{l}),\, \Delta_{y}(k_{t}, k_{l}) ).
\end{equation} 
These operators are essentially identical because they turn one into the other under exchange of the couplings 
$(k_{t} \to k_{l},\, k_{l} \to k_{t})$:
\begin{equation}
\Delta_{y}(k_{t}, k_{l})=\Delta_{x}(k_{l}, k_{t}),
\end{equation}
and for this reason we will focus our analysis on $\Delta_{x}.$
Let us thus write the equation for the component $x(n,t)$:
\begin{equation}
(\lambda- V_{n}) x(n) = \Gamma^{(in)}_{n} x(n-1) + \Gamma^{(out)}_{n} x(n+1)
\end{equation}  
\[
(\Gamma^{(out)}_{n} = \Gamma^{(in)}_{n+1})
\]
Here the coefficients  $\Gamma$ are c-numeric and have value $k_{l} (k_{t})$ at the horizontal (vertical) links;
correspondingly the site (vertex) potential is given by $V_{n} =-( \Gamma^{(in)}_{n} + \Gamma^{(out)}_{n}).$
By following a standard procedure,
this equation can be written in the form of a single step map acting on a two-dimensional space.
Once the transfer matrix $T \equiv T(n)$ has been assigned to the n-th vertex,
the vector $ \vec{z}(n) \equiv (x(n),x(n-1))^{\top} $ is mapped by $T$
into $\vec{z}(n+1)$: $\,\,\vec{z}(n+1)= T(n) \vec{z}(n).$ 
The transfer matrix $T \equiv T(n)$ has the form: 
\begin{equation}
T(n)_{1,1}= \frac{(\lambda - V_{n})}{\Gamma^{(out)}_{n}},\,\,\,\, T(n)_{1,2}= -\frac{ \Gamma^{(in)}_{n}}{\Gamma^{(out)}_{n}}
\end{equation}
\[
T(n)_{2,1}=1, \,\,\,T(n)_{2,2}=0
\]
In this way the eigenvalue problem translates into 
a boundedness condition for the  transfer matrix $T_{tot}$  of the total chain
$T_{tot} \equiv T(N_{k}) \cdot T(N_{k}-1) \cdot....\cdot T(2) \cdot T(1)$.
It is convenient to make use of $SL(2,R)$ (i.e. determinant +1) matrices,
because the boundedness condition for $T_{tot}$ can then be written as a single condition for the trace:  $\frac{1}{2} |Trace(T_{tot})| \le 1.$ \\
In view of that we perform the following transformation:
\[
T(n) \to \hat{T}(n) \equiv S^{-1}(n+1) \cdot T(n) \cdot S(n),\,\,\,\,S(n)=\hat{1}/(\Gamma^{(in)}_{n})^{1/2}
\]
In fact the new matrices $ \hat{T} \equiv \hat{T}_{out,in} =\hat{T}(\alpha, \beta)$ belong to $SL(2,R)$ and their entries are:
\begin{equation}
 \hat{T}_{1,1}= \frac{\lambda}{\beta} +\alpha+\frac{1}{\alpha},\,\,\,   \hat{T}_{1,2}=  -\alpha
\end{equation}
\[
 \hat{T}_{2,1}=  1/\alpha,\,\,\,  \hat{T}_{2,2}= 0
\]
The parameters $\alpha,\,\beta$ introduced above are functions of the hopping  coefficients:
$\alpha \equiv \alpha(n) =(\frac{\Gamma^{(in)}_{n}}{\Gamma^{(out)}_{n}})^{1/2}$ and 
$\beta \equiv \beta(n) = (\Gamma^{(out)}_{n} \cdot \Gamma^{(in)}_{n})^{1/2}.$
The iteration formula obtained in the Appendix (see Eq.A3), that we are going to represent by means of
transfer matrices, involves an operation $\mathcal{T}$ that describes the inversion of the ordering 
along the chain.  
When applied to a single step transfer matrix this inversion has the effect of exchanging $\Gamma^{(out)}$ with $ \Gamma^{(in)}$
because obviously it must be 
$\mathcal{T} (\hat{T}_{out,in}) = \hat{T}_{in,out}$; in terms
of the parameters $\alpha,\, \beta$ we have:
\begin{equation}
\mathcal{T}(\hat{T}(\alpha,\beta))=\hat{T}(1/\alpha,\beta) = \sigma_{1} \cdot \hat{T}^{-1}(\alpha,\beta) \cdot \sigma_{1},
\end{equation} 
where
 $\sigma_{1}$ is the Pauli matrix $[(\sigma_{1})_{12} = (\sigma_{1})_{21} = 1, \, (\sigma_{1})_{11} = (\sigma_{1})_{22}=0].$
The action of $\mathcal{T}$ on multiple step transfer matrices $M \in SL(2,R)$ is thus given by 
$\mathcal{T}(M) = \sigma_{1} \cdot (M)^{-1} \cdot \sigma_{1}.$ 
We label the transfer matrices
of the vertices with the directions of their links as follows:
$\hat{T}_{xy},\,\hat{T}_{xx}, \,\hat{T}_{yx},\,\hat{T}_{yy},$
where the left (right) label refers to the outgoing (ingoing) link.
The inflation rule for  $\hat{T}_{yx}$ and $\hat{T}_{yy}$ has the form:
\begin{equation}
\hat{T'}_{yx} =   \hat{T}_{yy} \cdot \hat{T}_{yx} \cdot \mathcal{T}(\hat{T}_{xy} \cdot \hat{T}_{yx})
\end{equation}
\[
\hat{T'}_{yy} = \hat{T}_{yx} \cdot \hat{T}_{xy} \cdot \mathcal{T}(\hat{T}_{yx} \cdot \hat{T}_{xy}),
\]
the 'conjugate' vertices
 $\hat{T}_{xy}$ and  $\hat{T}_{xx}$ transform according to the analog of this formula where the labels
$x$ and $y$ are exchanged.\\
By using the representation of $\mathcal{T}$ we have: 
\begin{equation}
\hat{T'}_{xy} =  \hat{T}_{yy} \cdot \hat{T}_{yx} \cdot \sigma_{1} \cdot \hat{T}^{-1}_{yx} \cdot \hat{T}^{-1}_{xy} \cdot \sigma_{1}
\end{equation}
\[
\hat{T'}_{yy} = \hat{T}_{yx} \cdot \hat{T}_{xy} \cdot \sigma_{1} \cdot \hat{T}^{-1}_{xy} \cdot \hat{T}^{-1}_{yx} \cdot \sigma_{1}.
\]  
From this equation we computed the band spectra on closed approximants of $PH_{2}$.
Since the chain pattern is invariant with respect to space reflexions,
it is sufficient to require that the solutions are bounded over a quarter of the chain.
Rather than  by diagonalizing products of exponentially
large numbers of matrices, we have studied the spectral problem in the space of the traces, as this approach allows for a much higher
precision. 
The boundedness condition quoted above translates into:
\begin{equation}
 \frac{1}{2} \cdot |Trace(\hat{T}_{yx} \cdot \hat{T}_{xy})| \le +1,\,\,\, \frac{1}{2} \cdot |Trace(\hat{T}_{yx})| \le +1.
\end{equation}
The relevant property to be stressed here is that the space of the traces is finite-dimensional. 
This is a direct consequence of the  
Cayley-Hamilton (CH) relation
\begin{equation}  
M^{2} - Trace(M) \cdot M + det(M) {\bf 1}=0.
\end{equation}
In fact it can be verified that (CH) allows to reduce the order of positive powers of matrices $M^{k},\, (k=2,3,...)$ 
and  to turn negative powers $M^{-1}$ into positive ones. 
More importantly one can show that the traceless parts $\mathcal{M}$ of the matrices $M$
\begin{equation} 
\mathcal{M} \equiv M - \frac{1}{2} \cdot Trace(M) \cdot {\bf 1}
\end{equation} 
satisfy  Clifford anticommutation rules.
This point can be easily ascertained by writing the $M$'s
in the real representation
\begin{equation}  
M= m_{0} \cdot \tau_{0} + \vec{m} \cdot \vec{\tau}, (\tau_{0} = {\bf 1},
\tau_{1} = i \cdot \sigma_{2},\, \tau_{2} =\sigma_{3},\, \tau_{3} = \sigma_{1}).
\end{equation}
This maps the matrices into a four-dimensional space, their coefficients being
\begin{equation}
m_{0} \equiv \frac{1}{2} Trace(M), \,\, m_{i}, (i=1,2,3).
\end{equation}
The algebra of the spin operators $\tau_{i},\, (i=1,2,3)$
allows to compute, for any couple of matrices $A,\,B$, the anticommutator  
$ [A, B ]_{+}$. By writing the result in terms of the traceless parts $\mathcal{A}$ and $\mathcal{B}$ one finally gets:
\begin{equation}
\mathcal{A} \cdot \mathcal{B}+ \mathcal{B} \cdot \mathcal{A}= (A;B) \cdot {\bf 1},
\end{equation}
where $(A;B) \equiv Trace(A \cdot B) - \frac{1}{2} \cdot Trace(A) \cdot Trace(B)$.\\
These rules allow to collect equal factors within products, so that their powers can 
be reduced to order one. 
In this way we get monomials were the four types of vertices are raised at the most to power one.
The occurrence of their $\mathcal{T}$-inverted partners introduces in the algebra
the matrix $\sigma_{1}$. We are thus led to operate in
the larger space  $SL^{\pm}(2,R)$. 
In the calculations we renamed the vertex matrices as follows:
\begin{equation}
 X(1) \equiv \hat{T}_{yx},\, X( \bar {1}) \equiv \hat{T}_{xy},\\
X(2) \equiv \hat{T}_{yy},\,  X(\bar{2}) \equiv \hat{T}_{xx}. 
\end{equation}
The elements $X(1),\, X(\bar{1})$ correspond to the orthogonal vertices; 
products where these factors occur in alternating order 
are associated with ladders aligned with the diagonals of the square lattice, as shown in Fig.4.
In general the transfer matrix of a chain with $n$ vertices has the form 
$X(\eta_{n},..,\eta_{l},...,\eta_{1}) \equiv X(\eta_{n}) \cdot..\cdot X(\eta_{l}) \cdot..\cdot X(\eta_{1})$ 
where the labels $\eta_{l}$ run over the four values $1,\, \bar{1}, \,2,\,\bar{2}.$
We wrote the trace map in terms of the following variables: 
\begin{equation}
x(\eta,\eta',....) \equiv Trace(X(\eta, \eta',..)),
\end{equation}
\[
y(\eta,\eta',...) \equiv Trace(X(\eta,\eta',....)\cdot \sigma_{1}).
\].
\section{\bf Multifractal spectrum} 
We will now illustrate the results obtained with the trace map for the scalar disordered laplacian $\Delta_{x}$.
We considered a very small anisotropy:
$k_{l} =1 - \epsilon,\, k_{t} = \frac{1}{k_{l}},\,\, (|\epsilon| <<1).$
When $\epsilon >0$ the x-variable at the horizontal links undergoes a restoring force
weaker than at the vertical ones. In other words its longitudinal displacements 
are slightly softer than the transversal ones.\\
One finds that
after one iteration the initial band  
is split into four bands. This branching process repeats itself at the successive steps  
so that  the number of bands $\mathcal{N}_{k}$ scales as the size of the system  $\mathcal{N}_{k} \approx 4^{k}.$\\
We have found that upon increasing the anisotropy the bands start to overlap.
The results reported here do not cover this case, as   
we made sure that the behavior $\mathcal{N}_{k} \approx 4^{k}$ was preserved up to $k=10$ iterations.\\
The spectra of 'disordered laplacians', when characterized by means of the density of states (DOS) $\rho(\lambda)$,
 can display a variety of scaling behaviors  $\rho(| \lambda |) \approx (| \lambda|)^{\alpha-1}$. 
Here instead of $\alpha$ we use the exponent $\mu$ that gives the scaling of the bandwidths $\Delta \lambda$
with respect to $\mathcal{N}$: $\Delta \lambda \approx \mathcal{N}^{-\mu}$. 
As illustrated in the caption of Fig.1 $\mu$ and $\alpha$ are simply related: $\mu = \frac{1}{\alpha}$.
In the literature on phononic spectra it is more usual
to consider, rather than $\alpha$ or $\mu$,
the spectral dimension $d_{s}$ \cite{rammal} that gives the scaling of the density of states with respect to the frequencies
$\rho(\omega) \approx (\omega)^{d_{s}-1}$. Since  
$\lambda \equiv - \omega^{2}$ the two exponents are trivially related: $\alpha =\frac{d_{s}}{2}$.
\begin{figure}
\centering
\includegraphics[width=0.8\textwidth]{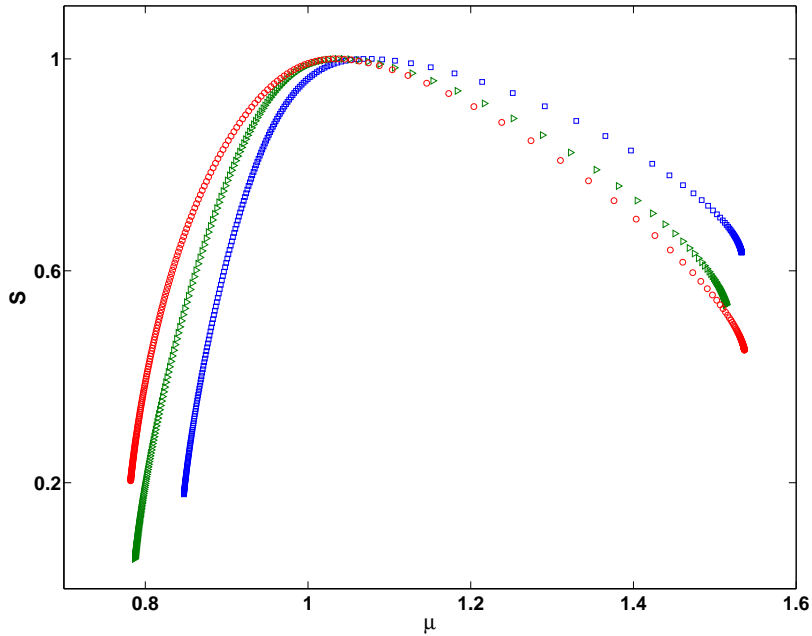}
\caption{Isotropic limit of the disordered laplacian.
We display here the distribution $S(\mu)$ of the scaling exponents $\mu$
obtained from the periodic approximant of order k=10.
The three  plots correspond to increasingly small values of the anisotropy parameter
($\epsilon = 10^{-5}$ red (circles), $10^{-6}$ green (triangles), $10^{-7}$ blue (squares).
It appears that the isotropic limit is characterized by a spectrum which turns out to be
qualitatively different from what would be expected in the case of periodic chains (see text).
We consider here the distribution for the
exponent $\mu$, which defines the scaling of the bandwidths 
$(\Delta \lambda \approx \mathcal{N}^{-\mu})$.
We recall that the more frequently used exponent 
$\alpha$ refers 
instead to the scaling of the density of states (DOS) 
$\rho(\lambda) \approx \lambda^{\alpha -1}$.
The two representations are equivalent; in fact 
from the boundedness of the integrated DOS 
   $H(\lambda):\Delta H(\lambda) \equiv \rho(\lambda) \cdot \Delta \lambda =O(1)$ 
one obtains  $\mu = \frac{1}{\alpha}$.
}
\label{Fig:1}
\end{figure}
Quite surprisingly we have found that close to the isotropic limit the spectrum has a whole set of power law singularities.
In other words the spectrum is strongly sensitive to disorder even with rather small values of $\epsilon.$ 
We did not sistematically explore the isotropic limit $\epsilon \to 0$
but the whole set of singularities \cite{luck1} found here is qualitatively different from what
would be expected with periodic chains. 
Here together with genuine normal modes, herefrom named Peano modes, corresponding to $\mu=1$,
we find singularities with powers below and above $\mu=1$.
\begin{figure}
\centering
\includegraphics[width=0.8\textwidth]{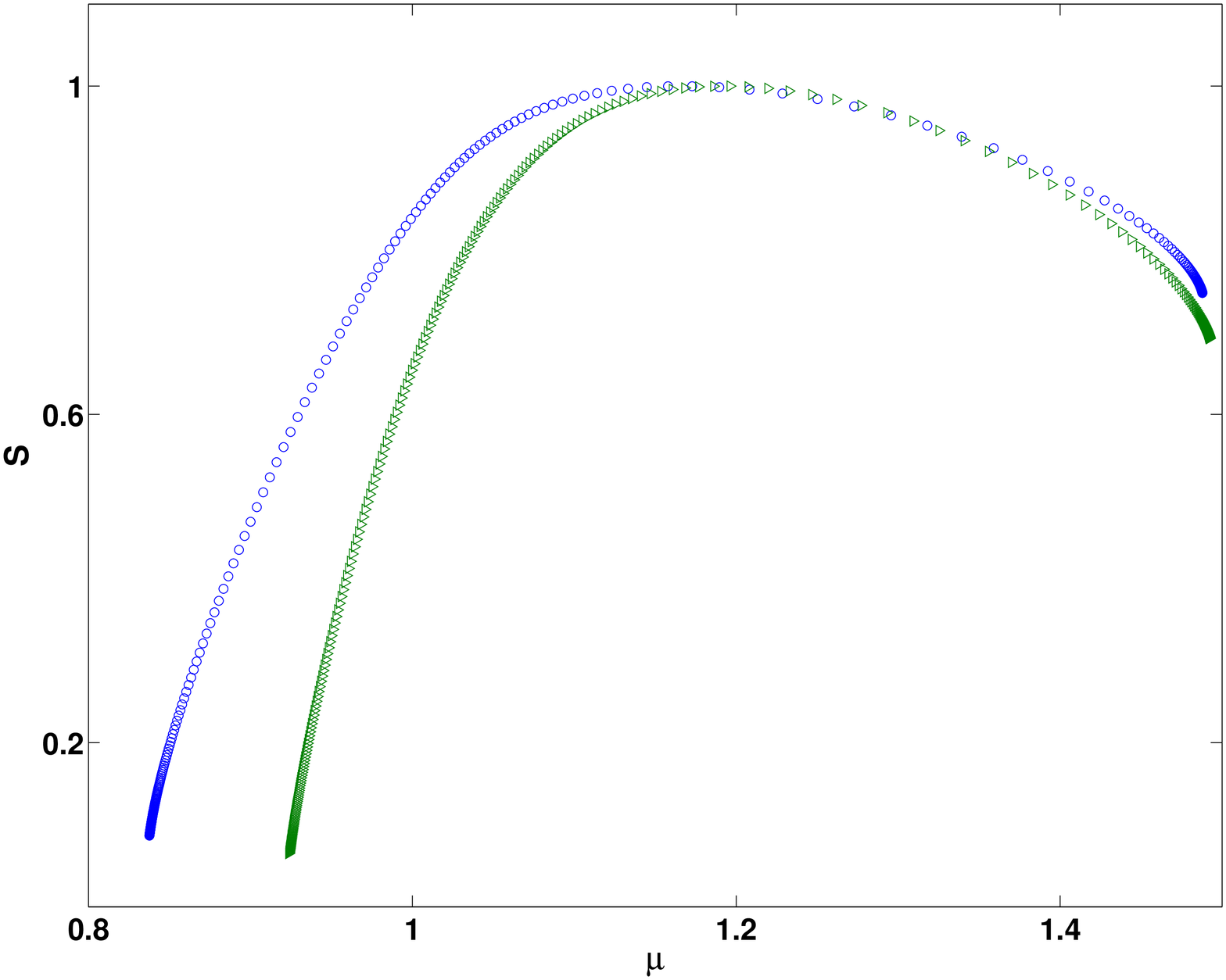}
\caption{Spectrum corresponding to the vector-valued eigenstates: plot of the distribution $S(\mu)$ 
obtained with inhomogeneity parameter $\epsilon = 10^{-4}$ (green, triangles).
For comparison we plot the distribution associated with the
corresponding single variable problem (blue,circles).
The interval $(\mu_{min},\mu_{max})$ appears to be contracted in the two-dimensional problem.
Thus the requirement that the eigenvalues 
are at the intersection of the two single-variable spectra $\lambda_{x}(k_{t},k_{l})$ and $\lambda_{y}(k_{t},k_{l})$
modifies the extremal regions $\mu \approx \mu_{min},\, \mu \approx \mu_{max}.$} 
\label{Fig:2}
\end{figure}
Working on discretized, disordered extensions
of laplacians such as the one examined here,
even within a small spectral range  $\Delta \lambda$ a variety 
of singularities can be found, so that the numerical plots of the DOS versus $\lambda$ are generally hard to 
interpret.   
This difficulty can be surmounted when
the singularities are of power-law type. In that case by collecting 
together the portions of the spectrum that share the scaling  exponent one can obtain a smooth 
distribution. 
This procedure goes under the name
of thermodynamic formalism  \cite{halsey} for multifractals.\\
One defines a partition function $Z(\tau)$, where $\tau$
is the analog of an inverse temperature; the sum 
is taken over the $\mathcal{N}$ bands of a periodic approximant, and accounts for their
individual widths $\Delta \lambda_{l}, \, (l=1,...,\mathcal{N})$:
\begin{equation}                              
Z(\tau) \equiv \Sigma^{\mathcal{N}}_{l=1} (\Delta \lambda_{l})^{-\tau} \approx (\mathcal{N})^{q(\tau)}.
\end{equation}
The exponent $q(\tau)$ is the analog of a free energy: in fact it is the sum of an entropy and an energy contribution.
To see this, let us consider  the  subset of bands $E(\mu)$ whose 
widths scale as $\Delta \lambda \approx \mathcal{N}^{-\mu}.$ If we make 
the dependence on $\mu$ explicit on the r.h.s. of the equation, we get:
\begin{equation}
q(\tau) = S(\mu) + \mu \cdot \tau,\,\, (\mu \equiv  \frac{d q}{d \tau}).
\end{equation}
Fig.1 is the plot of the 'entropy' function $S(\mu)$ that weights the different $\mu$-scalings within the spectrum,
so that its value corresponds to the fractal dimension of the subset $E(\mu)$.
The extremal values  
$\mu_{max,min} \equiv \lim_{\tau \to \pm \infty} \frac{dq}{d \tau}$ of the support of $S(\mu)$ characterize the 
'vacuum' (temperature T $\to 0_{+}$) 
and the 'fully inverted state' (temperature T $\to 0_{-}$) 
regimes. 
Notice that from the formula relating $q(\tau)$ with $S(\mu)$
one identifies the 'internal energy' U of the system:
\begin{equation}
 U \equiv - \frac{dq}{d \tau}
\end{equation}
So far we have considered the scalar operator $\Delta_{x}$, acting on the single variable x:
this problem describes a 'polarized' D=1 diffusion, as if a blind walker
experienced a sequence of varying mobilities encoding the path in its meandering along the plane. 
The results obtained for the full vector-valued problem  $\Delta_{(PH)}$  are shown in Fig.2, where they are also compared with
the scalar case. The two distributions turn out to be different, in particular the range of exponents $\mu$ is larger
in the scalar case.  
When the operator $\Delta_{(PH)}$ is diagonal in x and y,  
its eigenstates can be obtained by combining the solutions of the single variable problems in x and y:
\begin{equation}  
\vec{\alpha}_{\lambda}(n) \equiv (x_{\lambda_{x}}(n) , 0)^{\top}\\
\vec{\beta}_{\lambda}(n) \equiv (0 , y_{\lambda_{y}}(n))^{\top},
\end{equation} 
where $\lambda_{x}$ and $\lambda_{y}$ are the eigenvalues of
$\Delta_{x} \equiv \Delta_{x}(k_{t},k_{l})$ and  $\Delta_{y}   \equiv \Delta_{y}(k_{t},k_{l})$.
In order to have an eigenstate of $\Delta_{(PH)}$ with eigenvalue $\lambda$ it must be:
\begin{equation}
\lambda = \lambda_{y}(k_{t},k_{l})=\lambda_{x}(k_{t},k_{l}).
\end{equation}
In other words since the spectrum of $\Delta_{(PH)}$  is at the intersection of the two spectra
it is not surprising to find a smaller range of exponents.
\section{\bf Langevin equation}
We proceed now to examine the gaussian relaxation process associated with the disordered Laplacian 
$\Delta_{(PH)}$. This allows to understand the nature of the physical states. It turns out 
that the behavior of the fluctuations is determined by the spectral exponents $\mu$.
Our discussion relies on the literature dedicated to dispersion relations of fractals \cite{alexander} 
as applied e.g. to the Generalized Gaussian Structures \cite{gurtovenko}.
We assume that the vector field of deviations 
$\vec{X}(t) \equiv (\vec{x}(1,t), \vec{x}(2,t),....,\vec{x}(n,t),......,\vec{x}(4^{k},t))$   
evolves according to the following Langevin equation:
\begin{equation}
\frac{d}{dt}\vec{X}(t)= \mathcal{D} \cdot \Delta_{(PH)} \vec{X}(t)+\vec{\xi}(t)
\end{equation}
\[
<\xi_{i}(n,t)\xi_{j}(n',t')>=2 \cdot \mathcal{D} \cdot k_{B} T \cdot \delta_{i,j} \cdot \delta(t-t') \cdot \delta_{n,n'},\,
(i,j = x,y).
\]
This process can be considered as a generalization of the Rouse model, describing the
relaxation of an homogeneous polymer.   
The critical dynamics of self-avoiding Rouse chains  and membranes  has been studied by K.J.Wiese \cite{wiese}.
Here we are studying a purely linear system, but with a nontrivial dependence on the geometry
of the underlying pattern.
For the sake of simplicity let us omit from now on the vector indexes and 
put $D=1$ and  $k_{B} T =1.$
It is well known that
the process defined by a Langevin equation is fully
described by the response function (also called dynamic susceptibility)
and by the correlation function; 
these two functions are related through the fluctuation-dissipation theorem \cite{tauber}.
Before examining the case of the disordered laplacian
it is useful to remind some basics on gaussian processes.
We consider the relaxation of a field $\phi(x,t)$  in the presence of white noise $\xi(x,t)$: 
\begin{equation}
\frac{d}{dt}\phi(x,t)=-[r -\nabla^{2}]\phi(x,t)+\xi(x,t).
\end{equation} 
In the energy-momentum representation the dynamic susceptibility is 
$\chi(\lambda, \mathcal{Q}) = (- i \cdot \lambda +r+ \mathcal{Q}^{2})^{-1}.$\\   
The correlation function is defined as follows
$
C(x,t) \equiv <\phi(x,t) \cdot \phi(0,0)>=\int\frac{d \mathcal{Q}}{ 2 \pi} \int\frac{d \lambda}{ 2 \pi} C(\lambda,\mathcal{Q}) e^{i (\mathcal{Q} \cdot x - \lambda \cdot t)}$
and its energy-momentum form is:
\begin{equation}
 C(\lambda,\mathcal{Q}) =2 \cdot  (\lambda^{2} +  (r + \mathcal{Q}^{2})^{2})^{-1}.
\end{equation} 
We want now to relate the diffusive dispersion in the scale-free regime $r=0$ with the band-edge singularity in the spectrum
of the laplacian: more precisely we show that the $\mathcal{Q}^{2}$ dispersion is the counterpart of the 
multiplicity of states at the edges of the spectrum.
In order to make this clear we consider
the discretized version of the laplacian, where the counting of the states is elementary:
\begin{equation}
\frac{t}{2} (\psi_{n+1} + \psi_{n-1}- 2 \cdot \psi_{n}) = E \psi_{n},\\ 
(\psi(N)=\psi(0))
\end{equation}
Consider now the map $E \mapsto n$ associated with 
the set of eigenvalues
$ E=E(n) \equiv E(k_{n}) = t (cos(k_{n})-1),     (k_{n} = \frac{ 2 \pi n}{N})$: this map is
invertible with the exclusion of
the band edges   $E=0$ and $E= -2 t$, where
the DOS displays the D=1 van Hove singularities
\begin{equation} 
d\rho(E') = \frac{1}{2 \pi} \frac{dE'}{(t^{2}-E'^{2})^{\frac{1}{2}}},\, (E' \equiv E+t)].
\end{equation}
The points  $ n(E)$, when projected over the E-axis, produce in the interior of the band the scaling
$ |dE| \propto |dn|$ and $|dE| \propto |dn|^{2}$ at the edges. 
Since one has $|dn| \approx \mathcal{Q}$, the latter scaling 
is  the laplacian's dispersion $\mathcal{Q}^{2}$.\\
We generalize this connection to the case of the disordered laplacian: it is understood that our argument  
is merely heuristic because it refers only to the behavior in the scale-free regime,
that is to the region $\lambda \approx 0$ and $\mathcal {Q} \propto \frac{1}{N}$, $N$ being the chain length.
Recalling that within the spectral set $E(\mu)$ 
the singularity is $|\Delta \lambda| \approx \mathcal{N}^{-\mu}$, 
we associate to it a dispersion relation of the form
\begin{equation}
-\lambda(\mathcal{Q}) \equiv \omega^{2}(\mathcal{Q}) \approx |\mathcal{Q}|^{\mu}
\end{equation}
This amounts to saying that the contribution arising from $E(\mu)$ is described    
by an anomalous gaussian process  undergone by a field 
$\phi_{\mu}(\lambda, \mathcal{Q})$ with dynamical susceptibility
\begin{equation} 
\chi_{\mu}(\lambda, \mathcal{Q}) = (- i \cdot \lambda + |\mathcal{Q}|^{\mu})^{-1}.
\end{equation}
The correlation function
\begin{equation}
C_{\mu}(\lambda,\mathcal{Q}) =\frac {2}{(\lambda^{2} +  (|\mathcal{Q}|^{\mu})^{2})}
\end{equation}
when integrated over $\lambda$ and $\mathcal{Q}$, is represented as follows:
\begin{equation}
C_{\mu}(x,t) \equiv < \phi_{\mu}(x,t) \cdot \phi_{\mu}(0,0)> 
\propto  \int\frac{d\mathcal{Q}}{ 2 \pi}\frac{1}{|\mathcal{Q}|^{\mu}} exp^{i( \mathcal{Q} \cdot x - |\mathcal{Q}|^{\mu} \cdot t)}.
\end{equation}
The infrared behavior associated with $E(\mu)$ is thus:
\begin{equation}
C_{\mu}(n,0) \approx (n)^{\mu -1}, \,\, \, n>>1,\,\, (n(L)\approx L^{2}),\,\,C(0,t) \approx (t)^{1-\frac{1}{\mu}}, \, t>>1.
\end{equation}
Here we replaced the space variable $x$ with the chain's parameter $n$ and
$L$ indicates the linear size of the planar region covered by $n$ subsequent links.\\
In conclusion, having shown that the exponent $\mu$
determines both space and time behaviors in the infrared regime, 
we can classify the different types of fluctuations available to the system.\\
The region $\mu <1$ describes fluctuations that  decay in the length $n$ as a power law.
The corresponding eigenfunctions presumably have an oscillating behavior
where the amplitudes are progressively reduced with distance, as in destructive
interference. 
In addition, when $\mu <1$ the fluctuations are short-lived since they decay with time as a power law.\\
In the region $\mu >1$ we find the complementary behavior, where 
the fluctuations grow with $n$. They can be characterized in terms of
the fractal dimension $D_{f}=\frac{2}{\mu-1},$ or, in the polymeric interpretation, 
by means of the Flory exponent $\nu=\frac{1}{D_{f}} = (\mu - 1)/2.$\\
One can wonder how spatially growing functions could be acceptable eigenfunctions:
the point is that in the presence of oscillations
the boundedness condition can be recurrently fulfilled.
In this case there is presumably some sort of constructive interference.
The two regions $\mu<1$ and $\mu>1$ are complementary also with respect to time dependence:
in fact when $\mu>1$ the chain sites undergo subdiffusion
with an exponent  $\gamma_{D}= 1-\frac{1}{\mu}$. This behavior can as well be
characterized by the dynamic exponent $z= 2/\gamma_{D}$.
The scenario found here leads to associate $\mu >1$ and $\mu<1$ with excitations
respectively dominated by shear and by dilatation.\\
It is worth mentioning that  the fluctuation-dissipation theorem applies 
to the anomalous gaussian process described here.
As a consequence the well-known deGennes relation $z = 2 +D_{f}$ following 
from that theorem is fulfilled for all exponents $\mu$.
The case $\mu=1$, which marks the separation
of the two regions described above, corresponds
to oscillations with stable amplitudes.
The corresponding Peano modes will be discussed in connection with the symmetries of the system.\\ 
We conclude this Section with a comment on the general case of chains $PH_{D}$.
Being always generated by deterministic inflation, it is reasonable to expect 
that their self-similarity should be mirrored into
nontrivial spectral properties.
The eigenvalue problem in D dimensions, if approached along the lines presented in this paper,
would involve transfer matrices of dimension $(2D \times 2D).$\\
This can be difficult to deal with even in the case $D=2$:
we remind that it is possible to improve the numerical resolution, by using traces instead of matrices,
only when the components x and y are decoupled.
One can try to overcome this complexity by relying on a mean-field approach, which in the present context amounts  
to disregarding the fluctuations of  $\mu$.                                       
Taking into account that the chain is embedded in dimension $D$, self-consistency suggests 
that the mean-field exponent $\mu(D)$ should be chosen in such a way as to make the fractal dimension 
$D_{f}$ to coincide with the bulk's dimension: $D_{f}=D$. 
In order to determine $\mu(D)$ let us
consider a couple of  points (1,2) placed at a distance $r$ in the bulk, and use the formula 
$D_{f}=\frac{2}{\mu -1}$. In this way we find: 
\begin{equation}
<(\vec{x}(1)-\vec{x}(2))^{2}> \approx r^{2}= n^{\frac{2}{D}} = n^{\mu(D) -1},\,\mu(D)=1+\frac{2}{D}.
\end{equation}
Notice that in the language of membranes the fluctuating field $\vec{x}(n,t)$ undergoes the roughening transition 
at $\mu = 1,$  hence the mean field sets the membrane in the rough phase \cite{nelson}. \\
In the planar case we obtain $\mu(2)=2 \equiv D_{f}$.
It turns out that fluctuations of exponent $\mu=2$ would correspond to structures softer than 
the elastic network of $PH_{2}$ itself. In fact the maximal growth of the latter is determined by $\mu_{max}$ and we have      
found: $\mu_{max} < \mu(2) \equiv 2,\, \,(\gamma=0,\, \epsilon \to 0)$.\\   
In D=3 the mean field exponent is 
$\mu(3)= \frac{5}{3}$, which implies the anomalous diffusion $C(0,t) \approx t^{\frac{2}{5}}$.
This behavior was actually found to occur with high probability
in experiments on bacterial chromosomes \cite{weber,javer}
and is compatible with                                                  
configurations close to the optimal packing $D_{f}=3$  \cite{benza}.
Let us finally notice that fluctuations having
a fractal dimension higher than D cannot be excluded. In fact in some of the measurements quoted above  
\cite{weber} also rather small anomalous diffusion exponents $\gamma_{D}$ $(\gamma_{D} < \frac{2}{5})$ were obtained: 
they  would imply  $1 < \mu < \mu(3)$ and $ D_{f} > 3$.\\
\begin{figure}
\centering
\includegraphics[width=0.8\textwidth]{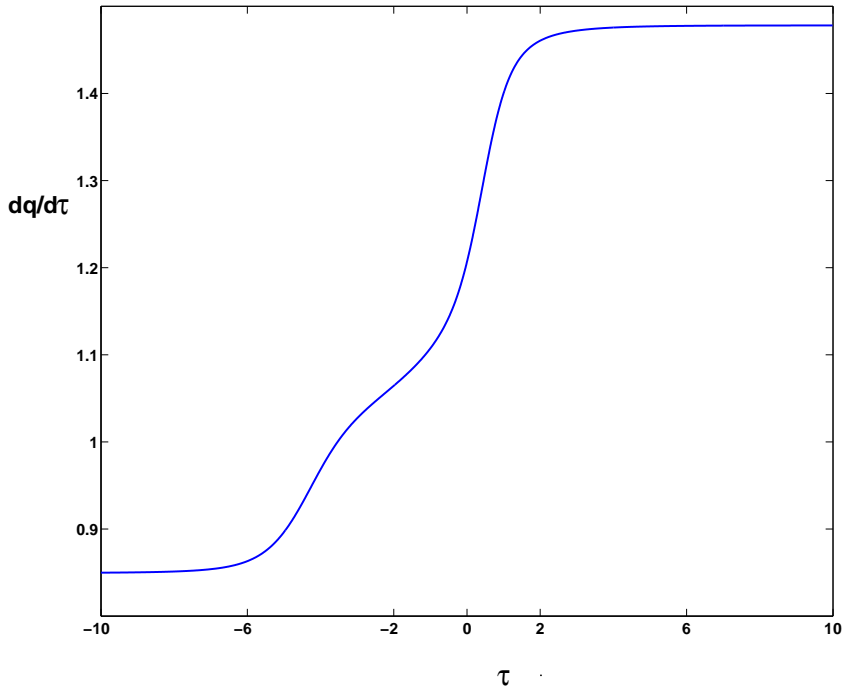}
\caption{Plot of the function $\frac{dq}{d\tau}(\tau)$ obtained from the bands of the approximant of order $k=10$.  
The inhomogeneity parameter is $\epsilon = 10^{-4}.$ The 'internal energy' U of the system
is given by $U = - \frac{dq}{d \tau}$. 
The asymptotic values of the function are respectively $\mu_{min}$, at large
negative values of $\tau,\, (\tau \to -\infty)$, and   $\mu_{max},\, (\tau \to +\infty)$ 
at large positive values of $\tau$.
At the critical (second order) point $\epsilon \to 0$ the function  $\frac{dq}{d\tau}(\tau)$  has a single inflexion 
at $\tau=0$.  The plot displayed here clearly shows three
inflection points, indicating the onset of the phase separation: the 'internal energy' $U$ is going to
develop two minima.
}
\label{Fig:3}
\end{figure}
\section{Inverse energy cascade and phase separation}
In this Section we describe the global properties of the system's relaxation
by  relying on the results discussed above.
We have found that the thermodynamic formalism of multifractals, which up to now  has been used primarily 
for technical reasons, involves a physical
interpretation of the 'internal energy' $U = -\frac{dq}{d\tau}(\tau)$ that
is consistent with the scenario of relaxation.
The region of negative $\tau$ describes the regime of negative temperatures (i.e. the inverted system)
and in fact, as one can see from Fig.3, there one finds the higher values of U. 
The limit $\tau \to +\infty$ corresponds to the ground state of the thermodynamics and there the 'internal energy' 
U has its minimum.\\
The interesting point, revealing some connection of U with the energies $\lambda$, is that   
the infrared region $\lambda \approx 0$  appears to be dominated by the largest scaling exponent $\mu_{max}> 1$:
at least this is what emerged from our numerical checks. 
As discussed in the previous Section the values $\mu >1$ are associated with excitations 
of a large size, presumably dominated by shear.
In the opposite case, i.e. whenever the dilatation dominates, one expects localized excitations and higher energies,
associated with values $\mu <1$.\\
If the 'internal energy' U is monotonic with respect to the energy content of the fluctuations, 
then from the plot reported in Fig.3 one gets a clear picture of the process of relaxation:
when prepared on localized states, the system is led to evolve towards the shear-dominated regime $\mu>1$. 
This behavior has been in fact observed in two-dimensional colloids \cite{yodh,kurchan}  and is intriguingly similar
to the inverse energy cascade of two-dimensional turbulence \cite{tabeling}.\\
It turns out that the analogy between this model and two-dimensional colloids is not limited to the space-time behavior
of the excitations, but involves also the spectral properties. In fact 
if one were to  add to the model the contact couplings among adjacent but nonsubsequent sites, 
the new network would describe a sort of textured membrane.
The resulting phononic spectrum, in addition to the contributions arising from the Peano pattern,
would display  in the density of states $\rho(\lambda)$ a contribution $\rho(\lambda)_{D=2} \approx \lambda$,
arising from planar phonons, as in the standard Debye spectra of two-dimensional systems. 
The quoted experiments on colloids did in fact display at low frequencies
an anomalously high density of states, superimposed to
the D=2 Debye behavior.\\  
So far we have discussed the spectral properties in the closeness of the isotropic limit, where the 
spectral bands are predominantly separated.
The question then arises about how the scenario changes when the anisotropy is large enough to make the bands overlap.\\
We have found that the anisotropy induces a phase transition \cite{artuso}: the 
spectral measure $S(\mu)$ progressively splits into two separate distributions.
Possibly the simplest example of a complete phase separation in a spectral measure is provided 
by the D=1 periodic chain. As we earlier recalled, in that case there are only two
exponents: $\mu=1$ and $\mu=2$. Since their weights are
$S(\mu=1)=1$ and $ S(\mu=2)=0$, the periodic chain is located at the extremal point of a line of phase coexistence,
at the extinction of the phase $\mu=2$, where the system is fully converted into the $\mu=1$ phase.\\
A phase coexistence was found \cite{calle} in the spectrum of the Quantum Ising Quasicrystal (QIQX).
The QIQX model describes a chain of spins where two Ising couplings are aligned in quasiperiodic order \cite{benzaq}.
When the two couplings are nearly equal $(J_{0} \approx J_{1})$
the function $S(\mu)$ has support  $\mu_{min} < \mu <\mu_{max}$ with $\mu_{min} =1$
and $\mu_{max}=2$. In other words the spectrum is at the coalescence of the two mentioned phases of the periodic chain. 
As the two couplings are made progressively different $(J_{0} \ne J_{1})$ the two phases separate.\\
In general a phase separation can be hardly detected by inspecting $S(\mu)$, because the numerics naturally converge to
the convex envelope of the two distributions. In fact $S(\mu)$ is the Legendre transform of the  object of relevance in this context, 
which is the 'internal energy' U $(U = -\frac{dq}{d\tau})$.
By closely inspecting Fig.3, where we display $\frac{dq}{d\tau}$ at the value $\epsilon = 10^{-4}$ of the anisotropy parameter,  
one can notice three inflection points: $A \approx (-4,.1.),\,B \approx (0.,1.1),\,C \approx (2.,1.3)$. 
At the (second order) critical point one expects a single inflexion point,
here we are very close to that point. Upon increasing the parameter $\epsilon$ 
a local maximum is expected in the region delimited by $(A,B)$. 
This maximum identifies the emergence on the left side of the plot, i.e. in the inverted regime, 
of a phase whose dominant scaling evolves towards $\mu_{min}$. 
On the right side the scaling of the other phase evolves towards $\mu_{max}$.
It is not difficult to associate the former with the solid, where the high energy dilatation modes are dominant,
and the latter, where the fluctuations grow both in $n$ and in $t$, with the shear-dominated liquid.\\ 
\begin{figure}
\centering 
\includegraphics[width=0.4\textwidth]{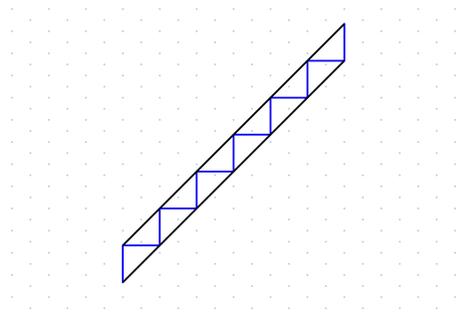}
\caption{Planar strip associated with products where the two types of orthogonal vertices $X(1),\, X(\tilde{1})$
are ordered in alternating fashion.
The horizontal (vertical) steps of the ladder correspond to discrete displacements 
along the space coordinates  $q_{x}, \, (q_{y})$. The ladder can be associated with a trajectory undergoing a sequence of scatterings 
at the strip's boundaries: the corresponding motion, aligned with the direction depicted in black,
would carry a linear momentum $p$ in that direction.
This suggests that the  orthogonal
vertices   $X(1),\, X(\tilde{1})$, which represent the turning points of the chain, should be associated with  
displacements in the plane of momenta $(p_{x},p_{y})$ where the axes form an angle $\pi/4$ with respect to the orientation
of $(q_{x},q_{y})$.} 
\label{Fig:4}
\end{figure}
\section{Critical crystal with eightfold symmetry}
In the previous Section we have shown that the model undergoes a phase transition,
and that its isotropic limit is found at the saddle separating two phases, respectively of solid and liquid type. 
The Peano network can thus be considered  a critical crystal, marking the point
where phase coexistence turns into coalescence. 
This situation appears to be the D=2 analog of the well studied critical regime at the D=1 delocalization transition 
\cite{aubryandre,harper}. It is thus appropriate to remind some results regarding that case.\\
The D=1 transition has been described by Wilkinson within a semiclassical (WKB) approach based  
on the universal hamiltonian \cite{wilkinson}:
\begin{equation}
\frac{1}{2} \cdot H \equiv cos(p)+ \alpha \cdot cos(q)
\end{equation} 
Exactly at the critical point $\alpha=1$ the hamiltonian has a twofold symmetry, 
since it is self-dual in the exchange of the (adimensional) space and momentum variables.
It identifies a square lattice in the $(q,p)$ plane, 
the lattice points being at the saddles where the separatrices meet.
These separatrices are soliton trajectories of energy $H=0$;
apart from them, every other orbit is closed, so that when $\alpha=1$ 
the particles are confined both in the $q$ and $p$ directions.\\
When the self-duality is broken, e.g. when $\alpha >1$, aligned solitons merge into trajectories running
parallel to the $p$-axis: in this region the particles are localized with respect to  the $q$-direction.
In correspondence with
the complementary coupling $(\frac{1}{\alpha})$ 
one finds the dual trajectories, where the  particles are  instead extended in the $q$-direction.
It resulted from Wilkinson's analysis that the critical properties obtained with the universal
hamiltonian, which has a twofold symmetry,  actually hold for an entire class of hamiltonians,
characterized by fourfold symmetry.
The following is an example of such hamiltonians:
\begin{equation}
\frac{1}{2} H \equiv cos(q)+cos(p)+cos(2\cdot q - p)+cos(q+ 2 \cdot p)
\end{equation}
We suggest that this fourfold symmetry could originate from the operators representing the 
discrete translations in  the $q$ and $p$ directions, together with their time-reversed
counterparts. Products of these four elements are identified by the oriented paths on the lattice phase space.
In order to see how the D=1 case is related with our construction it is
useful to examine the quantum hamiltonians based on chains organized according to the two-letter  
$(A,\,B)$ Fibonacci substitution rule.  
The Fibonacci chain is often depicted as a ladder over a two-dimensional lattice, where the
horizontal and vertical steps correspond to the letters $A$ and $B$.
The steps are combined in such a way as to fit the ladder within a strip 
having slope $m= \frac{1 + \sqrt{5}}{2}$. 
The ladder shown in Fig.4 is represented by products where the two operators $X(1)$ and $X(\bar{1})$,
associated with the orthogonal vertices of our construction, are ordered in alternating fashion.
In this elementary configuration the ladder is contained within a strip of slope $m=1$. 
One can notice that since the operators $X(1)$ and $X(\bar{1})$
involve couples of steps $(A B)$ or $(B A)$, they are more properly associated with translations
in the momenta along the diagonal. The quantum mechanics on the ladder has the form of a constrained D=2  hamiltonian system 
where translations in the momenta are reduced to a single dimension.\\
\begin{figure}
\centering
\includegraphics[width=0.4\textwidth]{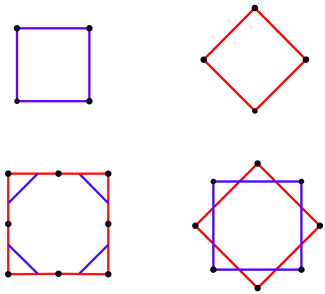}
\caption{Graphical representation of the elementary operations acting on the four-dimensional lattice.
Upper line: the blue (red) square path represents the loop of discrete translations in the p-plane (q-plane).
Lower line, right side: the eightfold motif obtained by projecting the $p$-loop on the plane of the $q$-loop.\\
Lower line, left side: the loop formed by combining the eight 
operators that generate the Peano chain. The colour convention is consistent with the rest of the Figure.
The red lines represent paths in the
q-plane, hence the red perimeter describes a loop in the q-plane. 
The black dots represent the vertices: the flat vertices $ X(2),\, X(\bar{2})$ carry only red lines,
the orthogonal vertices $X(1), \,X(\bar{1})$ are decorated with blue lines. 
The blue segments are aligned with the translations on the p-lattice, as illustrated in Fig.4.
If the blue segments are continued up to their intersections, the eightfold motif
shown on the right side is reproduced. 
} 
\label{Fig:5}
\end{figure}
The fact that at the critical point the D=1 quantum system is correctly described by the WKB approximation
translates, in the language of trace maps, into the condition
$Trace(A \cdot B \cdot A^{-1} \cdot B^{-1})=2$, meaning that the group commutator is the identity.
This condition is connected to the localization properties of the wave functions through 
the following trace relation:
\begin{equation}
Trace(A \cdot B \cdot A^{-1} \cdot B^{-1})= x^{2}+y^{2}+z^{2} - x y z -2
\end{equation} 
\[
x \equiv Trace(A),\, y \equiv Trace(B),\,z \equiv Trace(A \cdot B).
\]
On the r.h.s of the above one recognizes the well-known constant of motion of the trace map \cite{kohmoto}
representing the surface $I(x,y,z) \equiv x^{2} + y^{2} +z^{2} - x  y  z$.
The topology of this surface changes when $I=4$, i.e. when the group commutator is trivial, and
correspondingly two separate types of map trajectories (localized-delocalized)  meet.\\
We now return to the trace map of our case, leaving to future work \cite{bznext} a detailed analysis.
The map involves the monomials made out of the four vertices $(X(1), \, X(\bar{1}),\, X(2), \, X(\bar{2}))$.
One has six order-two monomials, four order-three monomials and one order-four monomial.
The degrees of freedom are doubled since the four basic vertices occur with their $\mathcal{T}$-reversed copies.
The map is thus defined over the traces $x = Trace(X)$ and $y=Trace(X \cdot \sigma_{1})$ illustrated in Section 2.
The overall space has dimension $15 X 2$ but this large number can be reduced by exploiting some symmetries.
For instance the invariance under cyclic permutations generates a triangular symmetry for cubic monomials. 
Cubic monomials are further related through some trace identities\cite{bznext}.\\
On top of that the occurrence of $4 \times 2$ basic vertices enforces an eightfold symmetry.
This, we believe, appears as the extension of the D=1 case that is characterized by a fourfold symmetry.
In fact in quantizing an hamiltonian system    
over a $(2 \times 2)$ phase space one needs  $(4 \times 2)$ operators 
representing the discrete $(q_{x},q_{y}),\,(p_{x},p_{y})$ translations and their reversals.
These operators act in this case on the vector of elastic displacements having values in a real two-dimensional 
vector space.\\
In order to understand how this geometry effects the configurations of the system, we project
the lattice of momenta on the plane of configurations.
In Fig.5 we enumerate the discrete translations of the phase space and compare them with the algebra of the vertices.
The motif displayed on the lower left is obtained by combining the vertices  $X(1),\, X(\bar{1}), X(2),\, X(\bar{2})$
with their reversed counterparts. 
On the lower right we show the effect of projecting the $(p_{x},p_{y})$ lattice over $(q_{x},q_{y})$. 
Notice that in the usual representation of the union of a q-lattice with its dual one assigns the nodes of the dual to the
centers of the q-plaquettes.
Under that convention the tiling of the plane would appear in the form of two equally oriented square lattices.
The eightfold motif obtained here results instead from the convention of making the centers of the  q and p plaquettes to coincide
and to having them rotated of an angle $\pi/4$.\\
We find that this convention reproduces the geometry associated with the operators
$X(1),\, X(\bar{1}), X(2),\, X(\bar{2})$. In fact the two motifs displayed on the second line 
give rise to identical periodic tilings of the plane.\\
\begin{figure}
\centering
\includegraphics[width=0.4\textwidth]{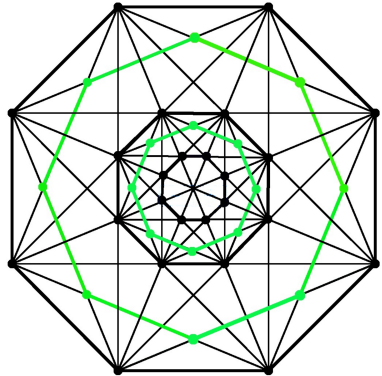}
\caption{The octagon and its descendants: this figure illustrates how the octagonal symmetry is compatible 
with dilatations. Discrete rotations and discrete dilatations match, as in regular crystals 
rotations match with lattice translations. This 'polar crystal' property is shared by the Peano chain
with the octagonal quasicrystal. 
Periodic approximants of the octagonal quasicrystal are obtained
by projecting the 4D hypercube on the plane; the resulting tiling turns out to have an average
connectivity substantially higher than in the Peano network. } 
\label{Fig:6}
\end{figure}
It turns out that in recent experiments on bilayer water, precursors of crystallization were observed to generate 
quadratic, hexagonal, pentagonal patterns, and more importantly, also the dodecagonal quasicrystal \cite{molinero}.
On very general grounds, since quasicrystals are much softer than solids, it is
somewhat to be expected that in the liquid to solid transition various structures of intermediate
stability should be explored by the system.\\
Given this eightfold symmetry the question is then: 
to what extent is the Peano network related with the octagonal quasicrystal?  
The effective coordination number of the network is certainly smaller than the average
coordination number of the octagonal quasicrystal.
Periodic approximants of the latter \cite{mosseri}, obtained by projecting the
4D hypercube on the plane, have square-shaped elementary cells.
Each cell has vertices with coordination numbers $z$ varying from three to eight $(z=3,4,5,6,7,8)$.
The free particle (laplacian) quantum motion defined over such tilings is characterized by anomalous diffusion, but with 
exponents much larger than those obtained here  \cite{benza2}. 
The reason for that can be easily understood, since
in the octagonal quasicrystal there is a larger number of channels open to hopping.\\
\begin{figure}
\centering
\includegraphics[width=0.8\textwidth]{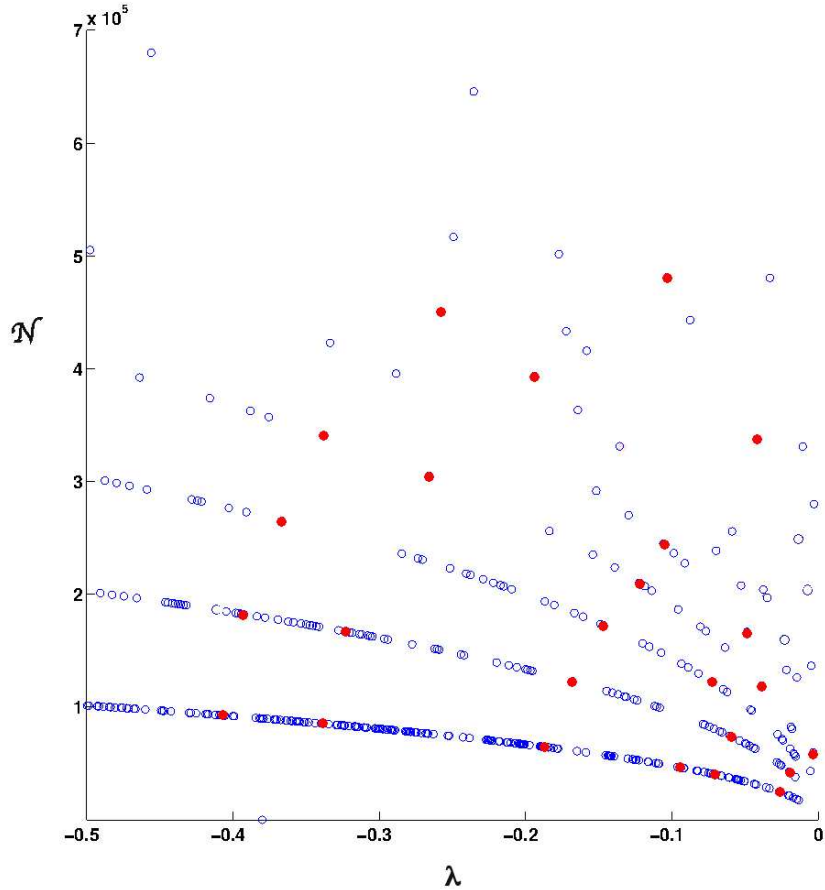}
\caption{Peano modes at the matching conditions of the 'polar crystal'.
Plot of the number of solutions $\mathcal{N}$ versus the energy $\lambda$ in the closeness of $\lambda=0$;
the distribution is defined over intervals having width $\delta \lambda = 10 ^{-8}.$  
Apart from the normalization, the plot is a coarse-graining  of the density of states (DOS) $\rho(\lambda)$.
The data refer to the approximant of order k=6 and to the anisotropy parameter $\epsilon= 10^{-6}$:
we are thus quite close to the critical point.
The points fall along five curves: $\mathcal{N}(l,\lambda),\,\, (l=1,2,3,4,5)$.
The curves $\sigma \equiv \sigma(\lambda,s)$ orthogonal to the $\mathcal{N}$'s are coloured in red.
The $\sigma$ curve intercepts approximants of increasingly large orders $l$.}
\label{Fig:7}
\end{figure}
We would like to stress that the octagonal symmetry arising in the Peano network
does not originate in the euclidean metrics, but in the hamiltonian structure.
If the system is integrable the hamiltonian can be made quadratic and in the isotropic limit the rotational
invariance implies the following form:
\begin{equation}
 H= \alpha \cdot (p_{x}^{2}+p_{y}^{2}) + \beta \cdot (q_{x}^{2}+q_{y}^{2}).
\end{equation}
At the delocalization transition the system is self-dual, thus at the critical point 
it must be $\alpha=\beta$:
\begin{equation}
(p_{x}^{2}+p_{y}^{2})= (q_{x}^{2}+q_{y}^{2}) = \frac{1}{2 \cdot \alpha} \cdot H
\end{equation}
The above means that the allowed states are points of the $(2 \times 2)$-dimensional lattice touched by the 
$S^{3}$ sphere of the energies and satisfying the self-duality condition.\\
\begin{figure}
\centering 
\includegraphics[width=1.\textwidth]{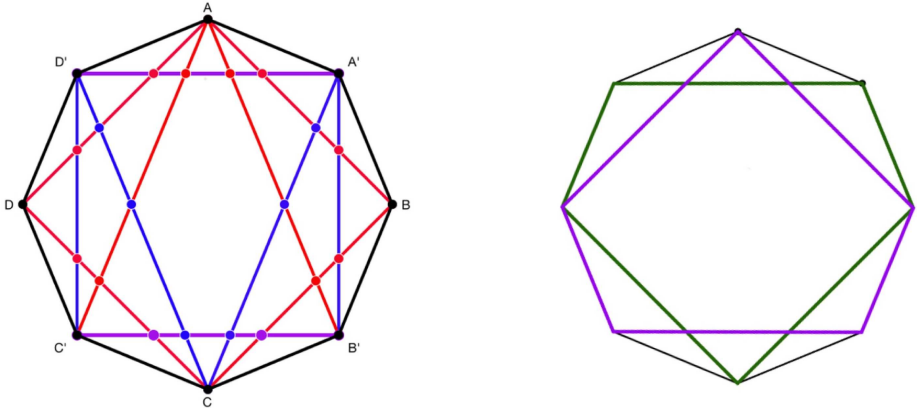}
\caption{Lower order symmetries compatible with the octagon. 
One can identify the fourfold loop in the $p$-plane (in blue) $A'B'C'D'$, and, in red, its analog in the $q$-plane
$ABCD$. Triangular loops mix the $q$ and $p$ coordinates. We show here two such loops: $AB'C'$ and $CD'A'$.
On the right side we display two fivefold loops  
obtained  by connecting the vertices left out of the triangular ones. The loop depicted in green is the complement of $AB'C'$.
Within a semiclassical description the loops can be associated with
scattering processes adding up to zero momentum exchange, as required by scale invariance at the critical point.} 
\label{Fig:8}
\end{figure}
This allows to understand why the Peano chain can sustain normal modes, in spite
of its intricate structure.
As shown in Fig.6 the octagon reproduces itself at discrete scales where the  dilatations match with discrete rotations. 
The chain shares with the octagon this 'polar crystal' property and can thus support
normal (Peano) modes, organized in a hyerarchical fashion. The numerical results 
confirm this argument. In fact by inspecting the multiplicity of solutions as a function of the energies $\lambda$ 
displayed in Fig.7 one can find some relevant features enumerated below.\\
A)The distribution appears as the union of several curves $\mathcal{N}(l,\lambda),\, (l=1,2,3,4,5)$ having shapes 
similar to what one would obtain with periodic chains. The curves can be associated with approximants of increasing order $l$.
In fact they represent solutions of  growing multiplicity, as one can see from their intercepts with the vertical axis.\\
B)A second family of curves $\sigma$ (in red) is apparently orthogonal to the $\mathcal{N}(l,\lambda)$.
The $\sigma$'s can be identified by means of their intercept $\lambda$ with the horizontal axis: 
$\sigma \equiv \sigma(\lambda,s),\,|\lambda \to \sigma(\lambda, s=0)$. As the arc parameter $s$ grows 
the curve $\sigma(\lambda,s)$ crosses approximants of increasing order $l$: 
the multiplicity (degeneracy) of the eigenvalue $\lambda$ increases accordingly.\\
The Peano waves can be depicted as Rayleigh waves whose packet is split into 
non-dispersive contributions corresponding to the different scales of approximants.
This scale separation actually involves not only the normal modes $(\mu=1)$,
but also the anomalous modes $\mu \ne 1$: one can identify the $l-$curves down to the 
edge $\lambda \approx 0$.\\
In addition to the octagonal symmetry, lower order ones, enumerated in Fig.8, can as well be of importance. 
The case of triangular symmetry was shortly illustrated above, in association with cubic monomials.
\section{\bf Conclusions}
We have studied the elastic deformations of the plane-filling closed Peano chain,
taken as an elastic network.
The shear and dilatation modes are intrinsically coupled, because their
waves are constrained to propagate along a thin planar strip undergoing the sequence of turns
dictated by the deterministic pattern of the chain.\\ 
The situation is thus similar to the interferences at the surface of an elastic material, 
giving rise to the dispersive Rayleigh waves.
The Peano chain has the property of being self-similar, with periodic approximants.
In a situation of this type the Rayleigh waves can find resonance conditions necessary
for coherent propagation. The corresponding normal modes, named Peano modes,
occur when the rescaling associated with the
inflation procedure matches with the discrete rotational symmetry of the pattern.\\    
We have studied the properties of the spectrum by means of the so-called thermodynamic formalism for multifractals.
In addition to the Peano modes we have found solutions that can be qualitatively described
as oscillating functions modulated by amplitudes which grow or decay: they are associated 
with gaussian relaxation processes characterized by a  scaling exponent $\mu \ne 1$.\\ 
These anomalous modes result from the frustrated coupling of shear and dilatation: when the
resonance conditions are not exactly fulfilled, the waves can be amplified or damped, depending
on how interference operates.
When $\mu <1$ the equal time correlation function decays along the chain 
as $n^{\mu -1}$, hence these states are of localized type; correspondingly the chain sites
undergo a damping process, with the power law $t^{1 -\frac{1}{\mu}}$.
The fluctuations characterized by $\mu>1$ have a complementary behavior: rather than damping they undergo anomalous diffusion
and their equal time correlation functions grow in amplitude along the chain.
They are predominantly found at low energies $(\lambda \approx 0)$. 
Taking the above into account, the relaxation  of the system can be described as a flow from
localized high energy structures towards extended low energy ones, where the latter survive
in the long time regime of the process. 
This  picture turns out to reproduce, at least qualitatively, the behavior observed in two-dimensional 
colloids \cite{yodh,kurchan}.
It may then be not coincidental that also the presence, observed in the same experiments,
of a low-energy bosonic peak can be readily explained within our model. 
In our opinion the peak  reveals
a contribution to elastic energy arising from closed, large-sized one-dimensional patterns.
Structures of this sort are not easily described within a membrane-like
network model. The density of states  should then account for these excitations, 
which add up  to the standard D=2 phononic part.\\
Possibly the main result obtained in this work is the proof of the existence of a critical point in the
isotropic limit of the model, where two phases coalesce.
These phases are respectively dominated by the $\mu <1$ and 
$\mu >1$ regimes: localized versus delocalized, or, in other words, solid versus liquid.
It is thus appropriate to consider the Peano network as a critical crystal.\\
Since, as previously mentioned, the dilatation modes have higher energies and are more localized, 
the energy flows towards extended configurations and lower energies, as in the inverse energy cascade 
of two-dimensional turbulence.
We have argued in addition that this critical point should be characterized by an eightfold symmetry: from
this perspective the Peano chain is closely related with the octagonal quasicrystal,
but has a much lower connectivity. 
Quite intriguingly, experiments on bilayer water revealed a transition from the liquid to the dodecagonal
quasicrystal\cite{molinero}.
The eightfold symmmetry that emerges here is strictly related with us 
limiting the description to contributions from large-sized D=1 structures; enlarging this picture could 
lead to results of a greater relevance for the phenomenology, 
but at the expense of losing the solvability.\\
From the perspective of the possible connections with the spatial ordering of chromosomes,
the  mean-field approach presented in Section 4 gives an estimate of the anomalous diffusion exponent
that fits quite well with the numbers observed in bacteria.\\ 
Let us finally comment on low energy excitations of a different kind that should as well be expected in Peano-shaped structures. 
The chain phonons treated in this work describe the small deviations with respect to a fully ordered pattern,
but in addition one could consider  isolated topological defects, giving rise to deviations 
from exact space-filling. 
The motion of such defects necessarily involves collective reorderings along the chain, i.e. phasons \cite{widom}.
We are led to associate with excitations of this kind the correlated motions  observed in chromatin \cite{zidovska}.
At even higher energies, multiple  defects should be accounted for, but this goes far beyond the perspective of our
approach; nonetheless, we point out that the chain phonons should give  a relevant contribution 
in the infrared regime, since they carry  the smaller momenta available to the system, scaling as $L^{-D}.$\\   
{\bf Acknowledgements}\\ 
The Author is indebted to M.Cosentino Lagomarsino for showing him that life may still be found
at the boundaries of maximally connected networks, and wishes to thank F.Prati and R.Artuso for reading
the paper.\\
\appendix
\section*{Appendix}
\setcounter{section}{1}
In this Appendix we determine the symbolic substitution rule allowing to generate
approximants of the closed Peano-Hilbert chain $PH_{2}$. To the best of our knowledge
the procedure illustrated here is original. It is organized in such a way as
to preserve the correct matching of the four quarters making up a closed chain,
as required in the spectral problem examined in the paper.
The approximant of order k of $PH_{2}$ is a path of length $N_{k} = 4^{k}$  that connects with no intersections
all points of the $(2^{k} \times 2^{k})$ square lattice. We display in Fig.A2
one of the mentioned quarters, corresponding to the approximant of order k=4.
Since the plane has the topology of $S^{2}$, a closed path cuts the plane in two polar regions.
At first sight it is equivalent to assign a model over the square lattice 
or over a Peano approximant, but the two choices are topologically different.
In fact the lattice is a torus while the Peano chain 
has the topology of a circle (see also the caption of Fig.A3).\\
Let us now enumerate all possible vertices, in order
to identify chain patterns with symbolic words.
We label the available vertices with the numerals $n$ and $\tilde{n},\, (n=1,2,..,6),$
where $n$ and $\tilde{n}$ are oppositely oriented as illustrated in Fig.A1.
\begin{figure}
\centering
\includegraphics[width=0.4\textwidth]{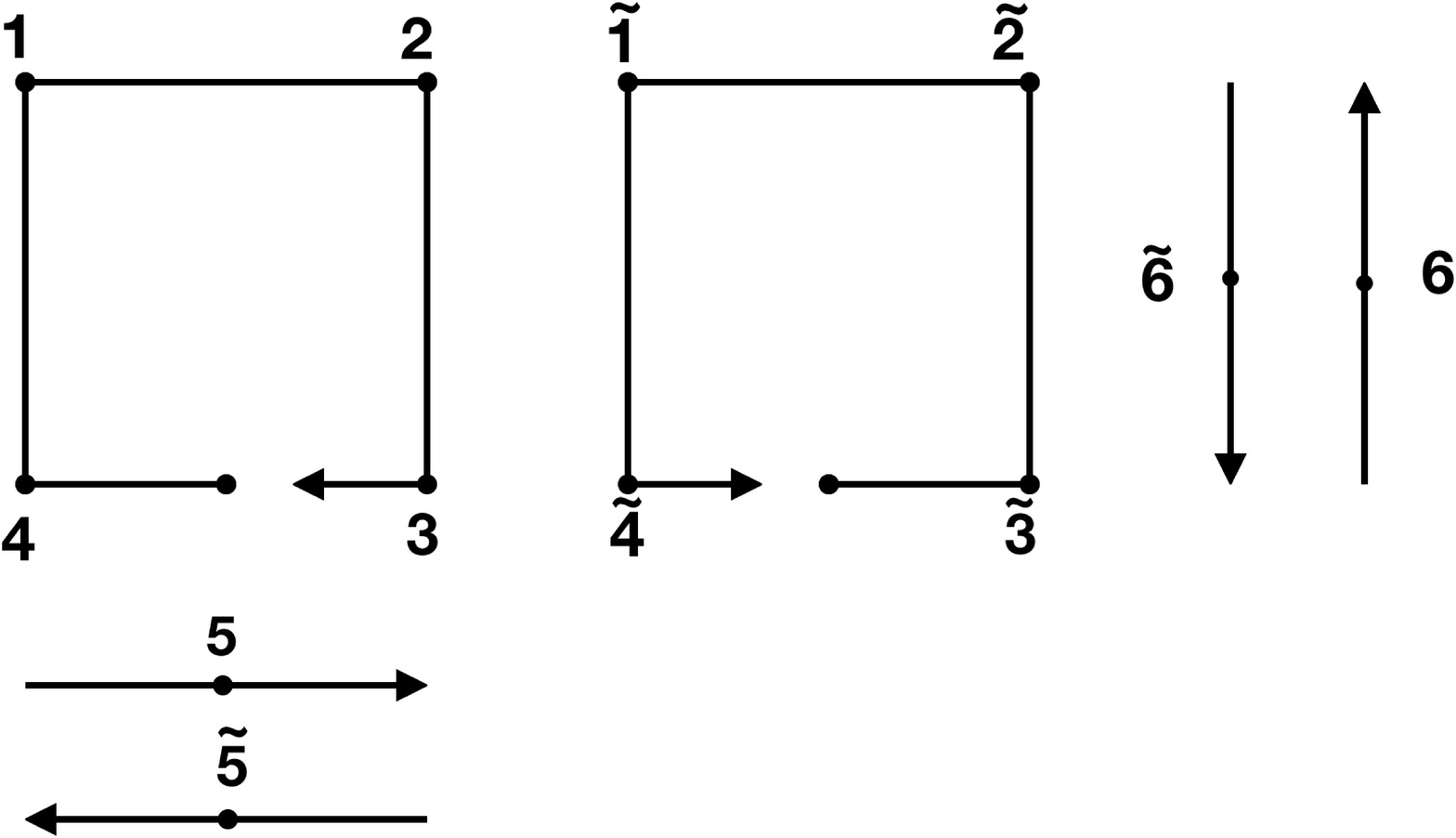}
\caption{The convention used for labeling the different types 
of vertices is here specified. As the iteration is based on a quadripartite structure, in order
to assign labels to vertices one can start with the elementary square-shaped
closed path. Since paths can as well go through lattice sites without bending, in addition to orthogonal
vertices also the flat ones (horizontal and vertical) must be accounted for.
With the numerals $(1,2,3,4)$ and $(5,6)$ we respectively classify
the patterns of orthogonal and flat vertices in the clockwise path; 
the vertices corresponding to the anticlockwise path are labeled with
 $(\tilde{1}, \tilde{2},\tilde{3}, \tilde{4})$ and 
$(\tilde{5},\tilde{6}).$}
\label{Fig:A1}
\end{figure}
The approximant of order $k$ is thus written as a symbolic word $W^{k}$ made out of the $6 X 2$ letters $l$ of
the alphabet defined above:
\begin{equation} 
W^{(k)} \equiv l(N_{k}) l(N_{k}-1).......l(3)l(2)l(1).
\end{equation}
The alphabet used for the vertices is exceedingly extended, as every vertex is the image of a flat or orthogonal
one under suitable transformations.
These transformations are the discrete rotation $R$,
the  inversions $P_{x},\,P_{y}$ with respect to the x and y axes  and the path inversion $\mathcal{T}$.
They act as listed here: 
$  \mathcal{T}(m) = \tilde{m},\,\mathcal{T}(m(n)....m(1)) \equiv \tilde{m}(1).....\tilde{m}(n)$;
$P_{x}(\tilde{4})=3,\, P_{y}(1)=\tilde{4},\, P_{y} (P_{x}(2))=4$;
$4=R(1), 1=R(2)$. We thus single out as basis elements the vertices $m=4$ and $m=6$.
\begin{figure}
\centering
\includegraphics[width=0.4\textwidth]{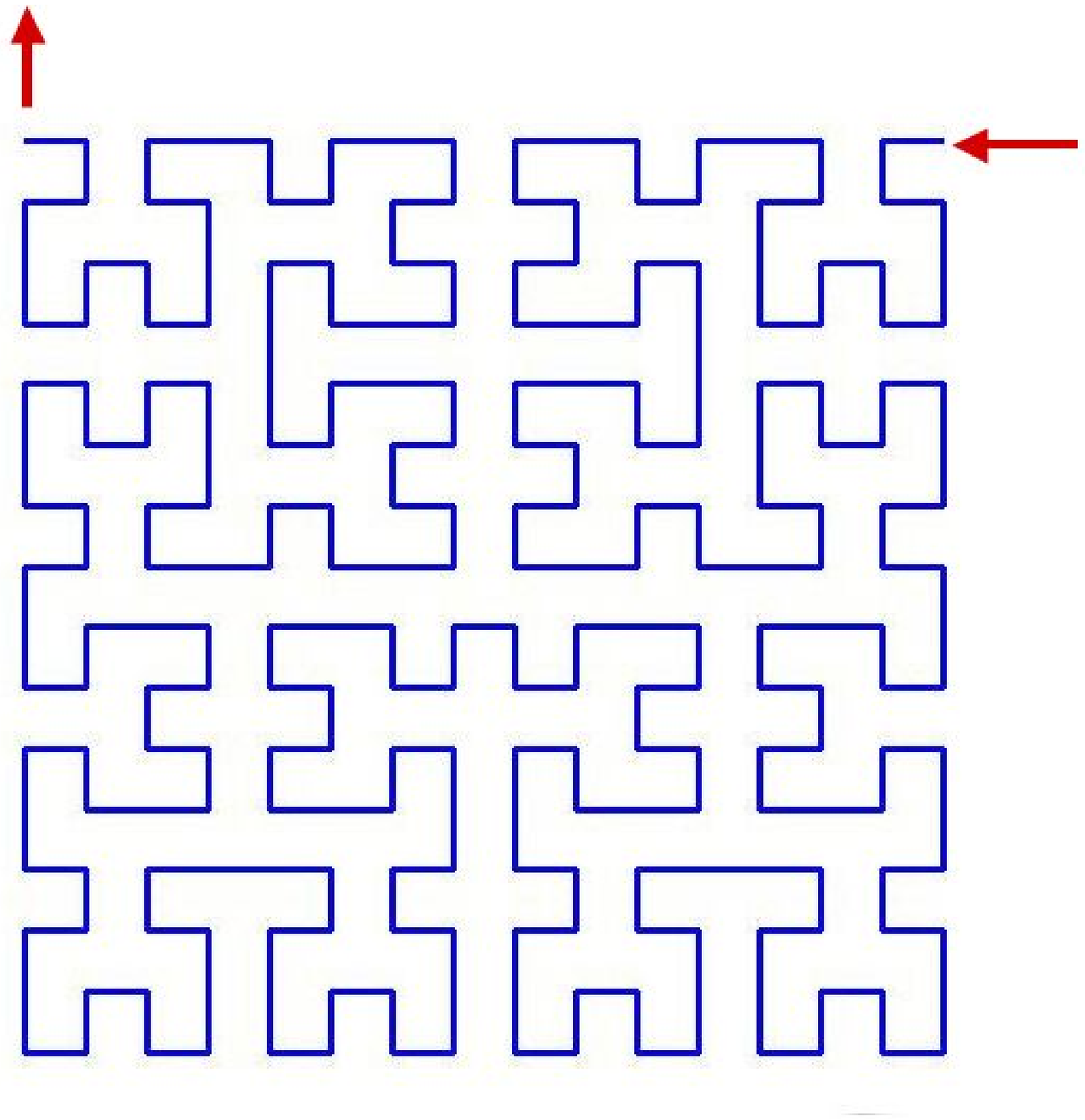}
\caption{Closed $PH_{2}$ chain, also known as Moore curve. The lower left quarter $(L\,\mathcal{L})$ 
of the approximant of order k=4 is displayed here.
The arrows depict the  ingoing and outgoing links connecting this quarter to its neighbours. The 
connection to the neighbour lying on the right side is provided by an horizontal link placed at the uppermost
available location.
A complementary situation regards the connection with the upper neighbour: in this case the link is
vertical and placed at the leftmost available location. The inflation procedure is defined in such a way as
to guarantee that this structure is preserved at all orders.  
} 
\label{Fig:A2}
\end{figure}
\begin{figure}
\centering
\includegraphics[width=0.4\textwidth]{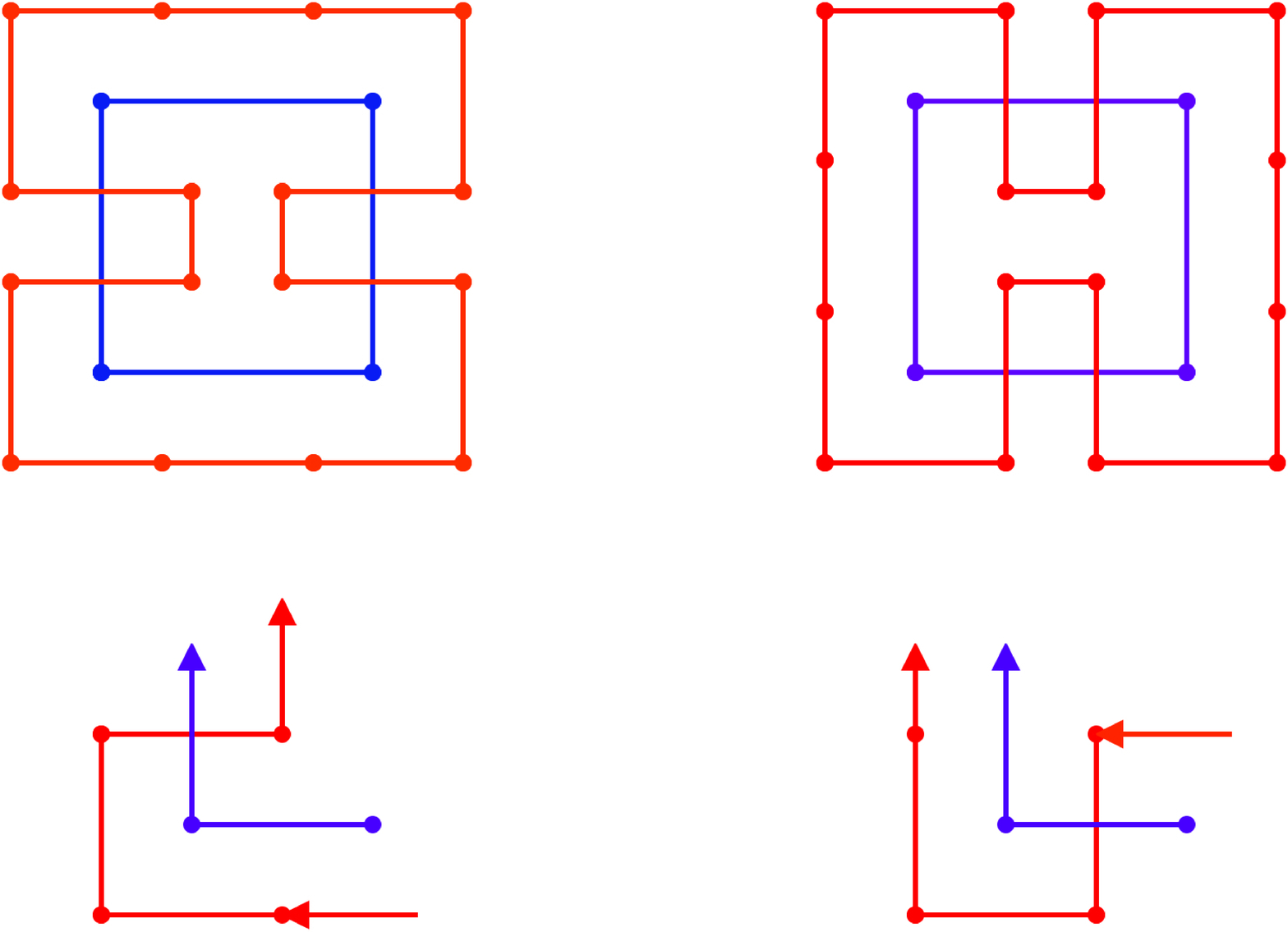}
\caption{In the lower line of this figure are displayed the two choices available 
when inflating the vertex $m=4$; the parent (inflated) patterns are depicted in blue (red).
With the left choice, the new vertices formed 
at the entrance and at the exit of the $m=4$ motif are respectively 
flat horizontal $(\tilde{5})$ and orthogonal 
$(\tilde{3})$. 
The convention chosed in the paper is displayed on the right: an horizontal link  enters 
the orthogonal vertex $(\tilde{1})$, and a vertical link exits the flat vertical
vertex $6$.
In the upper part of the figure we display the two resulting closed patterns after one step of the
inflation procedure; the initial configuration is the squared perimeter depicted in blue. 
The two complementary
H-shaped portions of the plane enclosed by the red patterns, when glued together as the two patches of a tennis ball, 
form a compact surface $S^{2}$. 
Hence the approximants separate the compactified plane in two polar regions.
The Peano chain is the maximally extended 'equator' of this partition. 
} 
\label{Fig:A3}
\end{figure}
Our aim is to write the substitution rule under the constraint of closed path.
Quite remarkably the constraint is sufficient to identify the rule.
Let us consider e.g. the choices available for inflating the vertex $m=4$:  
as shown in Fig.A3 two choices are available and they would imply twofold
branchings at each inflation step. This ambiguity is eliminated if one 
requires that the inflated path must be closed.
Every approximant can thus occur in only two forms; they are displayed
in the upper line of Fig.A3 for the first approximant. 
In our calculations we have chosen the option shown on the right column
of Fig.A3. 
With this choice the four quarters making up the chain are connected through the two external (left and right) vertical links and 
the two central horizontal links: the H-shaped pattern displayed on the right side of Fig.A3 reproduces
itself at all orders.   
Going back to the inflation rule, we write the convention
as follows: $4 \to 4' \equiv 4 \to (6 4 3 \tilde{1})$ 
and $6 \to 6' \equiv (4 \tilde{2} \tilde{3} 1).$
Considering the symmetries of the pattern, the following identities must hold at every order:
\begin{equation} 
 1=\mathcal{T} P_{y} (4), \,\,\, \tilde{3} = \mathcal{T} R (4),\,\,
\tilde{2} = R^{-1} P_{y} (4),\,\, \tilde{1} = \mathcal{T} R^{-1} (4),\,\, 3= \mathcal{T} P_{x} (4).
\end{equation}
We can thus write the inflation in terms of the two words
$Q_{4}$ and $Q_{6}$:
\begin{equation}
 Q^{(k)}_{6}=Q^{(k-1)}_{4} (R^{-1} P_{y} Q^{(k-1)}_{4})(\mathcal{T} R Q^{(k-1)}_{4})(\mathcal{T} P_{y} Q^{(k-1)}_{4})
\end{equation}
\[
 Q^{(k+1)}_{4}=Q^{(k)}_{6} Q^{(k)}_{4} (\mathcal{T} P_{x} Q^{(k)}_{4})(\mathcal{T} R^{-1} Q^{(k)}_{4}).
\]
The word $Q^{(k)}_{4}$ determines the lower left quarter of the closed chain $W^{(k)}$. 
The latter is obtained by operating with the symmetries and has the form:
$W^{(k)}= Q^{(k)}_{4} (\mathcal{T} P_{x} Q^{(k)}_{4})(P_{x} P_{y} Q^{(k)}_{4})(\mathcal{T}P_{y}Q_{4}^{(k)}.)$\\

{\bf Bibliography}\\
\bibliography{bib23july15}{}
\bibliographystyle{unsrt}
\end{document}